# Layer- and Gate-tunable Spin-Orbit Coupling in a High Mobility Few-Layer Semiconductor

Dmitry Shcherbakov[1], Petr Stepanov[1], Shahriar Memaran[2], Yaxian Wang[3], Yan Xin[2], Jiawei Yang[1], Kaya Wei[2], Ryan Baumbach[2], Wenkai Zheng[2], Kenji Watanabe[4], Takashi Taniguchi[5], Marc Bockrath[1], Dmitry Smirnov[2], Theo Siegrist[2], Wolfgang Windl[3], Luis Balicas[2], and Chun Ning Lau[1]*

[1] Department of Physics, The Ohio State University, Columbus, OH 43210
[2] National High Magnetic Field Laboratory, Tallahassee, FL 32310
[3] Department of Materials Science and Engineering, The Ohio State University, Columbus, OH 43210
[4] Research Center for Functional Materials, National Institute for Materials Science, 1-1 Namiki, Tsukuba, Ibaraki 305-0044, Japan
[5] International Center for Materials Nanoarchitectonics, National Institute for Materials Science, 1-1 Namiki, Tsukuba, Ibaraki 305-0044, Japan

**Spin-orbit coupling (SOC) is a relativistic effect, where an electron moving in an electric field experiences an effective magnetic field in its rest frame. In crystals without inversion symmetry, it lifts the spin degeneracy and leads to many magnetic, spintronic and topological phenomena and applications. In bulk materials, SOC strength is a constant that cannot be modified. Here we demonstrate SOC and intrinsic spin-splitting in atomically thin InSe, which can be modified over an unprecedentedly large range. From quantum oscillations, we establish that the SOC parameter $\alpha$ is thickness-dependent; it can be continuously modulated over a wide range by an out-of-plane electric field, achieving intrinsic spin splitting tunable between 0 and 20 meV. Surprisingly, $\alpha$ could be enhanced by an order of magnitude in some devices, suggesting that SOC can be further manipulated. Our work highlights the extraordinary tunability of SOC in 2D materials, which can be harnessed for *in operando* spintronic and topological devices and applications.**

## Introduction

In a crystal, the two-fold degeneracy of spins is protected by the combined inversion symmetry in both space and time. In the well-known Zeeman effect, an external magnetic field breaks the time reversal symmetry (TRS) and splits the spin degeneracy by $g\mu_B B$, where $g$ is the gyromagnetic ratio and $\mu_B$ is Bohr magneton. Alternatively, spin degeneracy can be lifted by spin-orbit coupling (SOC) when spatial inversion symmetry is broken, even in the absence of a TRS-breaking magnetic field. In crystals lacking structural inversion symmetry, the SOC coupling is of the Rashba form, $H_R = \alpha/\hbar \, (\hat{\mathbf{z}} \times \mathbf{p}) \cdot \boldsymbol{\sigma}$, where $\alpha$ is the Rashba parameter, $\mathbf{p}$ the momentum and $\boldsymbol{\sigma}$ the Pauli spin matrices(*2, 3*). As a result, the energy band of the crystal is spin-split; within each sub-band, the charges' spin is locked to their direction of momentum (Fig. 1a). A large SOC can be utilized to manipulate spins and generate interesting phenomena(*4-6*), such as the spin Hall effect, spin-orbit torque, and topological phenomena including the quantum spin Hall effect and quantum anomalous Hall effect.

---

* Email: lau.232@osu.edu

The advent of two-dimensional (2D) materials provides platforms with unprecedented opportunities for SOC tuning. For instance, significant SOC can be "endowed" to graphene by proximitization with transition metal dichalcogenides(*7-19*), and intrinsic SOC in 2D semiconductors has been studied by weak antilocalization measurements(*20-22*). Here we exploit the atomically thin sheets of InSe to demonstrate the extraordinary tunability of SOC in 2D materials.

InSe is a layered semiconductor with thickness dependent band gaps that range from 1 to 3 eV. Its direct band gap (except in monolayers), combined with extraordinarily high mobility (*23, 24*), makes it a very attractive electronic and optoelectronic material. The 2D unit cell of an InSe monolayer is composed of two vertically aligned In atoms, sandwiched between two planes of Se atoms. When projected onto a plane, an InSe monolayer forms a honeycomb lattice, where the metallic and chalcogen atoms occupy the two sub-lattices, respectively (Fig. 1b). Bulk crystals of InSe are found in several polytypes, including $\gamma$, $\beta$, $\varepsilon$, and $\delta$ phases(*25*). In this study, crystals are grown by the Bridgeman technique, and are confirmed by transmission electron microscope (TEM) to be predominantly $\gamma$–phase with rhombohedral stacking, though stacking faults and nanotwins are rather common(see Supplemental Materials). The $\gamma$–phase belongs to the space group R3m, and is non-centrosymmetric; this lack of inversion symmetry is a key ingredient that allows SOC-induced spin-split bands in the absence of an external magnetic field. Though SOC in InSe has been measured via weak localization studies on relatively thick flakes(20, 22), the values obtained have relatively large error bars, while the single-gated devices do not allow independent tuning of the Rashba SOC.

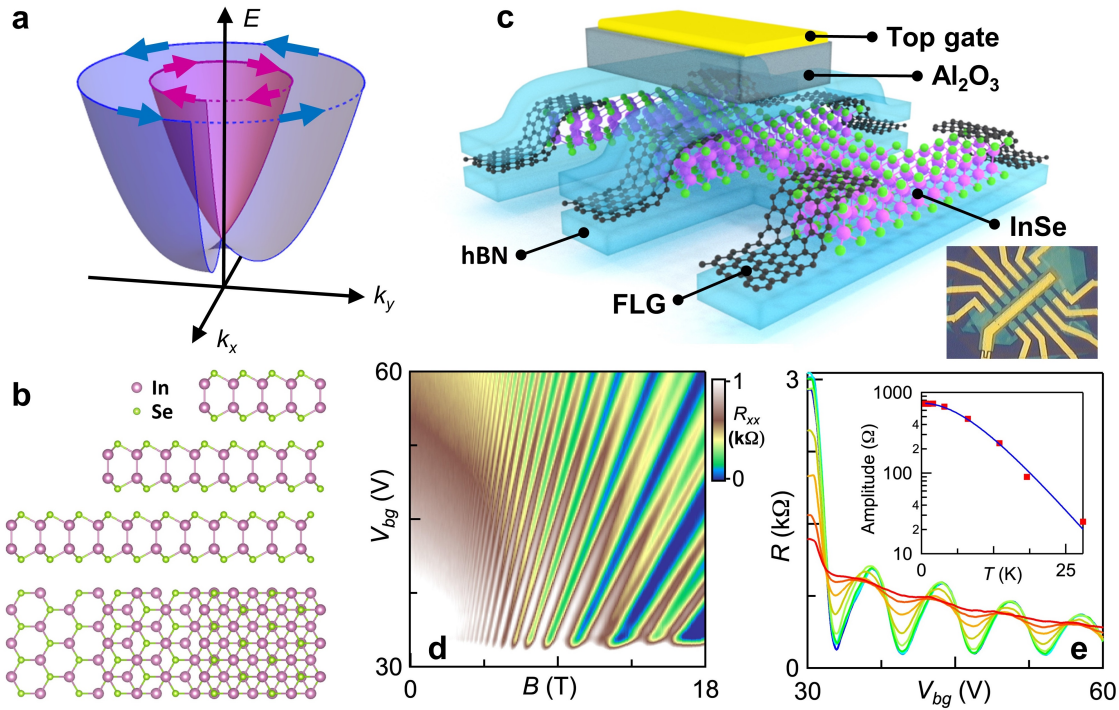

Fig. 1. Crystal structure, device schematics and magneto-transport data. (a). Rashba SOC-induced spin splitting of bands. (b). Side and top views of the crystalline structure of $\gamma$–InSe. The middle and bottom layers are partially exposed for better visualization (images created using VESTA software(*1*)). Inset: optical image of a device. (c). Schematics of device structure. (d). Landau fan $R_{xx}(V_{Bg},B)$ of device D1 from B=0 to 18 T at *T*=0.3K. (e). SdH oscillations at *B*=10T for temperatures ranging from 0.6K to 28K. Inset: Amplitude of the oscillations as a function of temperature. The line is a fit to the Lifshitz-Kosevich formula.

Our devices consist of atomically thin InSe sheets with thicknesses ranging from four to ten layers (Table 1), and sandwiched between hexagonal BN (hBN) layers, with few-layer graphene contacts, a silicon back gate and a metal top gate (Fig. 1c). By varying voltages applied to the top gate $V_{Tg}$ and back gate $V_{Bg}$, we can independently control the charge density $n = \frac{1}{e}(C_{Bg}V_{Bg}+C_{Tg}V_{Tg})$ and out-of-plane electric field $E_\perp = \frac{C_{Tg}V_{Tg}-C_{Bg}V_{Bg}}{2\varepsilon_0}$. Here $C_{Tg}$ and $C_{Bg}$ are capacitance per unit area between the gates and InSe, $e$ is the electron charge, and $\varepsilon_0$ is the permittivity of vacuum. We note that the application of $E_\perp$ is particularly crucial for modulation of Rashba SOC strength, which, depending on the direction of application, can either further break the inversion symmetry or compensate for the built-in asymmetry of the lattice.

**Results and Discussion**

Fig. 1d presents the longitudinal resistance $R_{xx}$ of device D1, which has a thickness ~6 layers, as a function of back gate voltage and magnetic field $B$. A prominent Landau fan is observed. Shubnikov de Haas (SdH) oscillations start at $B$ as low as 2.5T, indicating a quantum mobility exceeding 4000 cm²/Vs, which is the highest value reported to date. At $B$>12 T, quantized plateaus for filling factor $v$<10 are observed. To estimate the effective mass of the charge carriers, we extract the amplitude $A$ of the SdH oscillations at $B$=10 T at different temperatures (Fig. 1e). From the Lifshitz–Kosevich equation, $A=CT/\sinh(bT)$, where $T$ is the temperature, $b=\frac{2\pi^2 k_B T}{\hbar \omega_c}$, $\omega_c = \frac{eB}{m^* m_0}$ the cyclotron frequency, $m^*$ the reduced effective mass, $m_0$ is the rest mass of electrons, $k_B$ the Boltzmann constant, $\hbar$ the reduced Planck constant, and $C$ is a constant, we estimate that $m^*$~0.14, in excellent agreement with prior reports(23, 24). Similar measurements on 4-layer and 10-layer devices yield $m^*$=0.15 and 0.13, respectively. Thus we will use $m^*$=0.14 throughout the rest of the manuscript.

Table 1. Device Parameters

| Device | Estimated Thickness | $\alpha_0$ (10⁻¹¹ eV m) | $\alpha'$ (e nm²) |
|---|---|---|---|
| D1 | 6-layer | 3.5 | 3.0 |
| D2 | 4-layer | 0.6 | -1.2 |
| D3 | 7-layer | 0.2 | -1.2 |
| D3 | 8-layer | 0.15 | -1.15 |
| D3 | 9-layer | 0.08 | -1.1 |
| D4 | 6-layer | 0.3 | 1.1 |
| D5 | 4-layer | 0.8 | 1.3 |

The high device mobility enables manifestation of the hallmark of Rashba SOC – in 2D conductors with spin-split bands, the different concentrations of spin up and spin down electrons give rise to a characteristic beating pattern in the SdH oscillations. Fig. 2a displays $dR_{xx}/dB$ for device D2, which is estimated to be ~4-layer thick, as a function of $n$ and $B$ at a constant $E_\perp$=-0.33 V/nm, where the nodes in the oscillations appear as a white band for $B$~4 to 6T. A line cut of $dR_{xx}/dB(B)$ at $n$=7x10¹² cm⁻² is shown in Fig. 2b, where the nodes are indicated by the arrows. The presence of two distinct frequencies is clearly reflected in the Fourier transform (FT) of the data (Fig. 2c), where we observe two distinct peaks in $1/B$, and the separation between the two peaks is larger at higher charge density.

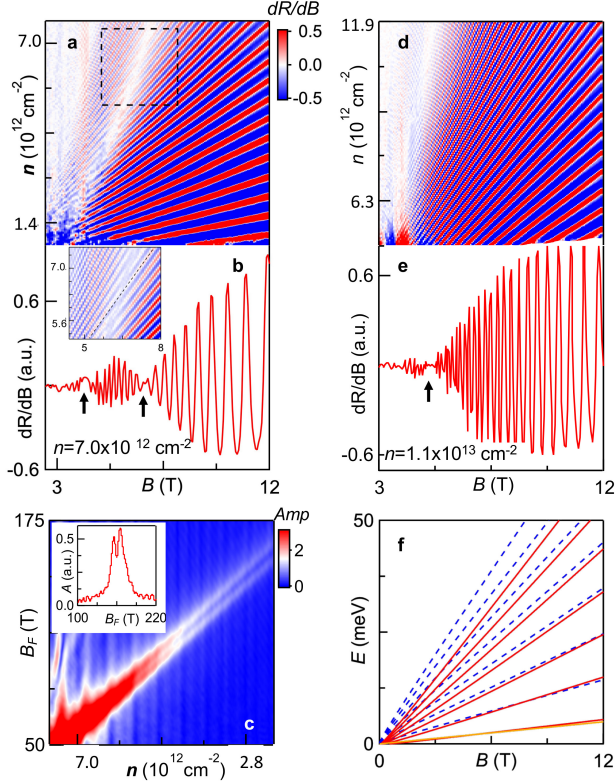

Fig. 2. Beating patterns in SdH oscillations from device D2 at $T$=0.3K. (a). $dR_{xx}/dB(n, B)$ at $E_\perp$ =0.33 V/nm. (b). Line trace of (a) at $n$=7.0x10$^{12}$ cm$^{-2}$. Arrows indicate nodes of oscillations. Inset: a zoom-in plot of the dashed square in (a). Dotted line indicates the transition from minima to maxima. (c). Fast Fourier transform of the data in (a), display two distinct frequencies. Inset: line trace of FFT amplitude vs $B_F$ at $n$=2.8x10$^{12}$ cm$^{-2}$. (d). $dR_{xx}/dB(n, B)$ at $E_\perp$=0 (e). Line trace of (d) at $n$=1.1x10$^{13}$ cm$^{-2}$. (f). LL energies calculated using Eq. (1) and $\alpha$=1.0x10$^{-11}$ eVm. The orange line denote the zeroth LL, whereas the red solid and blue dotted lines represent spin-up and spin-down levels for $N>0$.

Intriguingly, the parity of the oscillations changes after crossing the nodes – as shown in Fig. 2b inset, as $B$ increases, the conductance minima at a fixed filling factor become maxima and vice versa. Since the degeneracy of the oscillations is two at these fields, such a change in parity indicates that only odd integer states are resolved for $B$ below the nodes, and even integer states for $B$ above the nodes.

The beating pattern, the beating frequency that scales linearly with charge density and the alternating parities in the SdH oscillations are the hallmark signatures of large Rashba SOC in 2D semiconductor, in which the quadratic band is spin-split by $\Delta_R=2|\alpha k_F|$ (here $k_F$ is the Fermi momentum). The different Fermi surface areas of spin-up and spin-down bands give rise to different oscillation frequencies, which interfere to generate the characteristic beating pattern. The Landau level (LL) spectrum is then given by(2, 3, 6)

$$E_N^\pm = \hbar\omega_c \left[N \pm \tfrac{1}{2}\sqrt{(1-gm^*/2)^2 + N\frac{\Delta_R^2}{E_F \hbar\omega_c}}\right], \quad E_0 = \tfrac{1}{2}\hbar\omega_c \qquad , \qquad (1)$$

Here $N$ is the LL index, $E_F$ the Fermi energy, and $g$ the effective g-factor. Fig. 2f plots the LL spectrum for $N$=0 to 10. Significant LL crossings occur between spin-split bands. Notably, the Rashba SOC strength can be extracted from the beating patterns,

$$\alpha \approx \frac{\hbar^2}{m^* m_0}\sqrt{\frac{\pi}{2}}\frac{\Delta n}{\sqrt{n}}, \qquad (2)$$

where $\Delta n=(e/h)B_{F,beat}$ is the difference in densities between majority and minority spin carriers, and $B_{F,beat}$ is the beating frequency in SdH oscillations. Using Eq. (2), we estimate that $\alpha$~0.9x10$^{-11}$ eV m for the data set shown in Fig. 2a-c.

Since Rashba SOC is a consequence of broken inversion symmetry, we expect that an externally applied $E_\perp$ can either induce additional asymmetry, or compensate for the inversion symmetry of the lattice, thus providing an experimental knob to tune the SOC strength *in situ*. This is borne out experimentally – a similar data set of device D2 taken at $E_\perp=0$ (Fig. 2d) display a substantially different beating pattern: comparing with that at $E_\perp=-0.33$ V/nm, the nodes shift to smaller *B*, indicating that the SOC strength is reduced.

To systematically examine the variation of SOC with $E_\perp$, we measure $R_{xx}$ of device D2 as a function of *n* and $E_\perp$ at constant *B*=6T (Fig. 3a). As $E_\perp$ varies, the oscillations' maxima clearly transition into minima and vice versa; the transition points moves approximately linearly on the *n*-$E_\perp$ plane. In the SdH oscillations regime, the magnetoresistance of the device is related to the LL spectrum by(*26*)

$$\rho_{xx} \sim B^2 \sigma_{xx} \sim B^2 \sum (N \pm \tfrac{1}{2}) e^{-(E_F - E_N^{\pm})^2/\Gamma^2} \quad (3)$$

where $E_F$ is the Fermi energy, and $\Gamma$ is the LL broadening. We model the effect of the electric field by taking

$$\alpha = |\alpha_0 + \alpha' E_\perp| \quad (4)$$

where is the "intrinsic" Rashba parameter, arising from the asymmetry of the intrinsic lattice and/or the confining potentials, including the different dielectric thicknesses for the gates; $\alpha'$ parametrizes the effectiveness of the electric field at tuning the SOC coupling, which can be positive or negative, depending on the field's orientation relative to the built-in inversion asymmetry of the lattice. Combining Eq. (1) and (3), and assuming $\Gamma$=1 meV, $m^*$=0.14, we are able to satisfactorily reproduce the data by using $\alpha = 0.6 - 1.2\, E_\perp$, where $\alpha$ is in unit of $10^{-11}$ eVm and $E_\perp$ in V/nm (Fig. 3b). Magnitude of these values agree with those obtained from DFT calculations (Fig. 4d) (*27, 28*). Taken together, these results underscore the highly tunable nature of the Rashba SOC in a 2D semiconductor, which, by applying an external $E_\perp$, can be enhanced, reduced and even completely suppressed to zero.

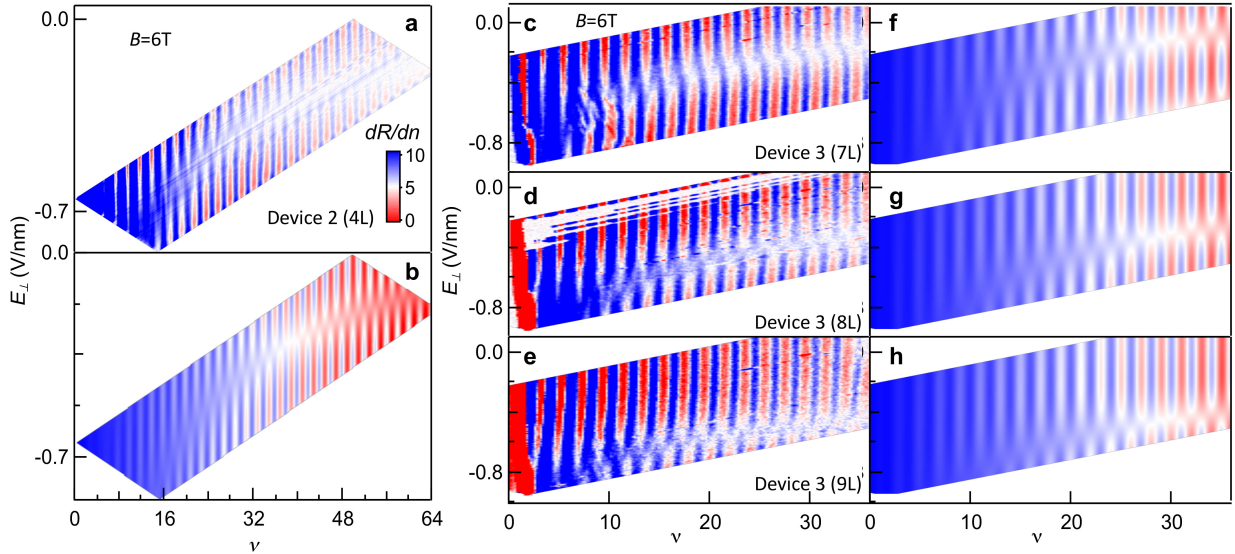

Fig. 3. Magneto-transport data $dR_{xx}/dn(n, E_\perp)$ at constant *B*. (a-b). Data and simulation for device D2. (c, d, e). Data for different regions of device D3 that are 7, 8 and 9 layers thick, respectively. (f, g, h). Simulations for device D3.

A unique opportunity afforded by the atomically thin 2D membranes is the exploration of SOC in a layer-by-layer fashion. To this end, we fabricate devices on a *single* piece of InSe flake with adjacent regions that differ by a single atomic layer in thickness[25]. Fig. 3c-e presents data for such a device D3, which hosts regions that are estimated to be 7, 8 and 9 layers, respectively. The overall features in the $dR_{xx}/dn(n, E_\perp)$ plots at $B$=6T are qualitatively similar, with an important difference: as the thicknesses increases, the transition points shift to lower (more negative) $E_\perp$ values. These patterns can be satisfactorily captured by our simulations, yielding the intrinsic Rashba parameter $\alpha_0$=0.2, 0.15 and 0.08x$10^{-11}$ eVm, respectively, for the regions that are 7, 8, and 9 layers thick; $\alpha'$, the effective tunability of the SOC, is extracted to be -1.2, -1.15 and -1.1 x 10 e nm$^2$, respectively. Thus, as the layer number increases, both the intrinsic SOC strength and its scaling with external $E_\perp$ are diminished(*27*); surprisingly, this trend is opposite to that predicted by theory calculations(*28*).

Finally, we observe that in some devices appears to be dramatically enhanced. Fig. 4a presents $dR_{xx}/dn(n, E_\perp)$ at $B$=10 T for device D1. Strikingly, instead of a single transition point at a given $n$, several crossing points are visible. Such large number of transitions indicates much higher SOC – indeed, we are able to reproduce the patterns by taking $\alpha = 3.5 + 3.0\, E_\perp$. Compare to values in previous plots, increases by as much as an order of magnitude, while tunability is enhanced by a factor of 3. Such a large SOC strength is unexpected, and unlikely to originate from variations such as thicknesses of flake or gate dielectrics. Instead, it suggests a fundamental mechanism at play. One possibility is variations in the interlayer spacing $c$, which, according to DFT calculations, is an extremely effective "knob" to tune the SOC parameter. For instance, increasing $c$ by a mere 1.9% (from 8.43 to 8.6 Å) results in a reduction of $\alpha_0$ by 57% (from 0.89 to 0.57 x $10^{-11}$ eVm) (Fig. 4d). The small variations in interlayer spacing may arise from stacking faults: even though the $\gamma$-phase is the most dominant phase in our crystals, TEM studies reveal high densities of stacking faults that are generated by dislocation dissociation or by growth, including ones with wrongly stacked monolayers (see Fig. S1-2 of Supplementary Materials). The large electrostatic pressure generated by the displacement fields, which exceeds 20 MPa in our devices, may further compress the interlayer spacing. Another possibility is the non-linear scaling of $\alpha$ with $E_\perp$ in the presence of stacking faults or twin boundaries. Though further work is required to ascertain the mechanism, this data show that Rashba SOC can be tuned through a much larger range than previously thought possible.

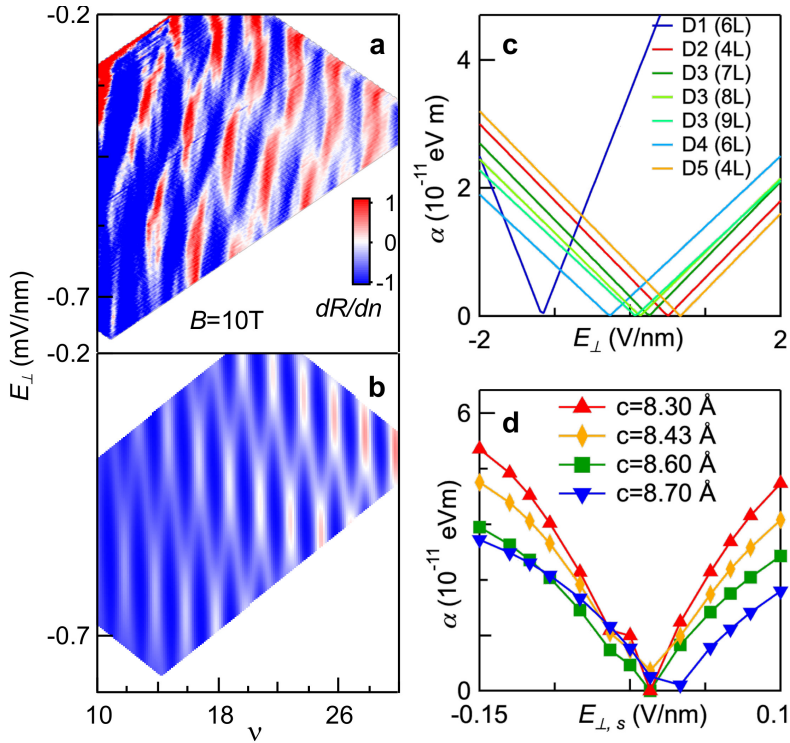

Fig. 4. Magneto-transport data and anomalously large SOC parameter. (a-b). $dR_{xx}/dn(n, E_\perp)$ and simulations at constant $B$=10T for device D1. (c). Extracted Rashba parameter and its dependence on $E_\perp$ for different devices. (d). DFT calculations of Rashba parameter vs $E_{\perp,s}$ for different c-axis lattice constants. Here $E_{\perp,s}$ is the out-of-plane electric field used in calculations, which, due to screening, is significantly reduced from that calculated from the experimental gate voltages.

In summary, we observe the hallmarks of large Rashba SOC in few-layer InSe field effect transistor devices, where the unprecedented mobility enables quantitative extraction of the Rashba parameter. The two-dimensionality of the devices enables the SOC strength to be modulated by the layer thickness, and by an electric field that breaks or compensates for the inversion symmetry. The possibility of piezo-SOC tuning, which has not been observed to date, warrants further theoretical and experimental investigation. The extraordinary tunability of SOC in 2D materials, as demonstrated here, can be exploited for a variety of phenomenon and devices, such as spin Hall effect, spin-orbit torque, and spin helix with long spin coherence time, with in situ control.

**Acknowledgement**
We thank Vladmir Fal'ko for stimulating discussion. The experiments are supported by NSF DMR 1807928 and NSF-DMR 1807969. MB acknowledges the support of DOE BES DE-SC0020187. Y.W. and W.W. acknowledge primary funding by AFOSR project no. FA9550-18-1-0335. Computations were performed on the machines of the Ohio Supercomputer Center under project no. PAS0072. K.W. and T.T. acknowledge support from the Elemental Strategy Initiative conducted by the MEXT, Japan, Grant Number JPMXP0112101001, JSPS KAKENHI Grant Numbers JP20H00354 and the CREST(JPMJCR15F3) JST. A portion of this work was performed at the National High Magnetic Field Laboratory, which is supported by National Science Foundation Cooperative Agreement No. DMR-1644779, and the State of Florida.

**Author Contributions**
D. Sh, P.S. and C.N.L. conceived the project. D.Sh. fabricated samples and performed measurements. P.S. assisted with sample fabrication. J.Y. and D. Sm. assisted with measurements. D.Sh., M.B. and C.N.L. analyzed and interpreted the data. S.M., W.Z., R.B. and L.B. synthesized InSe crystals. Y.W. and W.W. performed DFT calculations. Y.X., K-Y.W., R.B., T.S performed crystallographic measurements and analysis. K.W. and T.T. provided hBN crystals. D. Sh. and C.N.L. wrote the manuscript. All authors discussed and commented on the manuscript.

**Competing interests:** The authors declare that they have no competing interests.

**Data and materials availability:** All data needed to evaluate the conclusions in the paper are present in the paper and/or the Supplementary Materials.

**Materials and Methods**

Single crystals of InSe were synthesized from 6N-pure indium and 6N-pure selenium pellets in a ratio $In_{0.52}Se_{0.48}$. The starting materials were sealed in a quartz ampule and then vacuumed to $10^{-3}$ Torr. They were subsequently pre-reacted by gradually heating the ampule to

800 °C and kept at this temperature for 96 h. After the reaction, the ampule was then placed into an RF furnace where the RF power was gradually increased to raise the temperature up to 800 °C. The ampule was then pulled through the hottest zone at a rate of 2 mm/h.

Bulk InSe crystals are characterized by atomic resolution high-angle-annular-dark field scanning transmission electron microscopy (HAADF-STEM) using a probe-aberration corrected JEOL JEM-ARM200cF at 200kV. TEM samples were made by focused ion beam in a Helios G4 DualBeam™.

InSe and hexagonal BN (hBN) crystals are exfoliated on polydimethylsiloxane (PDMS) into atomically thin sheets, with InSe thicknesses ranging from 4 to 10 layers. Few layer graphene (FLG) for contacts is exfoliated on Si chips covered with 300 nm $SiO_2$. Using the dry transfer technique(29) at room temperature, we assemble hexagonal hBN/few-layer InSe/hBN heterostructures, with contacts that consist of few FLG sheets. First, we pick up the FLG contacts, followed by InSe and, finally, the bottom hBN. The exfoliation and stack assembly processes are performed in a glove box to minimize degradation. We then etch the heterostructure into the Hall bar geometry in two steps first using $SF_6$ gas for hBN, then Ar gas for InSe (pressure 15 mTorr, power 60 W, average etch time 15 s and 5 min for $SF_6$ and Ar correspondingly), and deposit a layer of aluminum oxide covering the channel, followed by Cr/Au for electrodes and the top gate. The electrodes formed one-dimensional contact to FLG.

Transport measurements are performed using SR830 and SR860 lock-in amplifiers at an ac bias current of 50 nA. All measurements are perform at base temperature in $^3$He or pumped $^4$He cryostats.

We used *ab initio* density functional theory (DFT) to investigate the thickness dependence of functional properties in atomically thin InSe. DFT calculations were performed using the Vienna ab initio Simulation Package (VASP)(30, 31) within the generalized gradient approximation (GGA) with Perdew-Burke-Ernzerhof (PBE)(32, 33) functional. A plane-wave cut-off of 350 eV and Γ-point centered $12 \times 12 \times 1$ $k$ mesh for integration in reciprocal space are employed(34). While we keep in-plane periodic boundary conditions, a vacuum of 12 Å is added perpendicular to the plane to minimize interactions between the periodic images of the films. Grimme's D2(35) functional is employed in all calculations to include the van der Waals interaction between the layers. When applying an external field on the slab, we turned on the dipole correction along the *c* axis to avoid interactions between the periodically repeated images. In order to determine the Rashba splitting for the bottom of the conduction band, fully relativistic spin-orbit coupling (SOC) was included into the calculations in addition to the applied field. In order to determine the position of the shifted conduction band minima accurately, band energies were calculated for a large number of k-points in the vicinity of the Γ-point, and the Rashba parameter (α) is determined by $\alpha = \frac{\Delta E}{2\Delta k}$.

Table 1. Device Parameters

| Device | Estimated Thickness | $\alpha_0$ ($10^{-11}$ eV m) | $\alpha'$ (e nm$^2$) |
|---|---|---|---|
| D1 | 6-layer | 3.5 | 3.0 |
| D2 | 4-layer | 0.6 | -1.2 |
| D3 | 7-layer | 0.2 | -1.2 |
| D3 | 8-layer | 0.15 | -1.15 |
| D3 | 9-layer | 0.08 | -1.1 |
| D4 | 6-layer | 0.3 | 1.1 |
| D5 | 4-layer | 0.8 | 1.3 |

Fig. 1. Crystal structure, device schematics and magneto-transport data. (a). Rashba SOC-induced spin splitting of bands. (b). Side and top views of the crystalline structure of $\gamma$–InSe. The middle and bottom layers are partially exposed for better visualization (images created using VESTA software(*1*)). Inset: optical image of a device. (c). Schematics of device structure. (d). Landau fan $R_{xx}(V_{Bg},B)$ of device D1 from B=0 to 18 T. (e). SdH oscillations at $B$=10T for temperatures ranging from 0.6K to 28K. Inset: Amplitude of the oscillations as a function of temperature. The line is a fit to the Lifshitz-Kosevich formula.

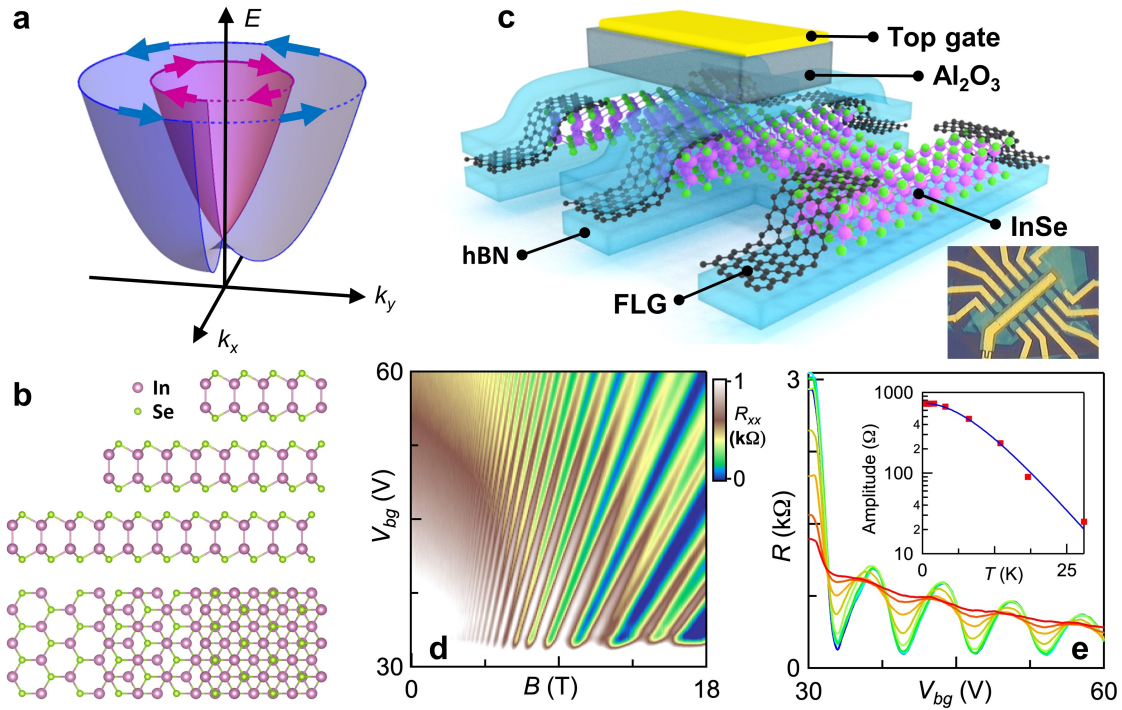

Fig. 2. Beating patterns in SdH oscillations from device D2 at $T=0.3$K. (a). $dR_{xx}/dB(n, B)$ at $E_\perp=0.33$ V/nm. (b). Line trace of (a) at $n=7.0\times10^{12}$ cm$^{-2}$. Arrows indicate nodes of oscillations. Inset: a zoom-in plot of the dashed square in (a). Dotted line indicates the transition from minima to maxima. (c). Fast Fourier transform of the data in (a), display two distinct frequencies. Inset: line trace of FFT amplitude vs $B_F$ at $n=2.8\times10^{12}$ cm$^{-2}$. (d). $dR_{xx}/dB(n, B)$ at $E_\perp=0$ (e). Line trace of (d) at $n=1.1\times10^{13}$ cm$^{-2}$. (f). LL energies calculated using Eq. (1) and $\alpha=1.0\times10^{-11}$ eVm. The orange line denote the zeroth LL, whereas the red solid and blue dotted lines represent spin-up and spin-down levels for $N>0$.

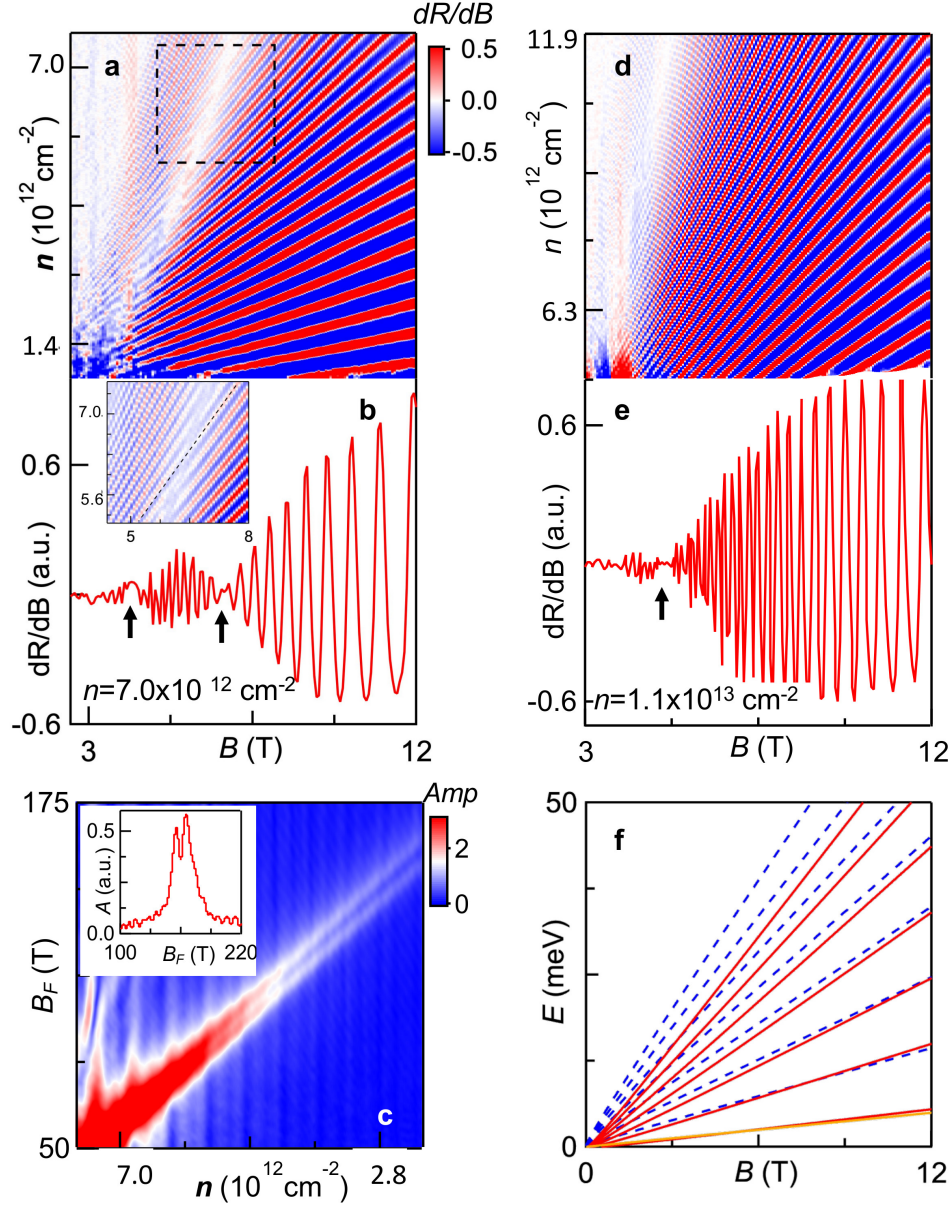

Fig. 3. Magneto-transport data $dR_{xx}/dn(n, E_\perp)$ at constant $B$. (a-b). Data and simulation for device D2. (c, d, e). Data for different regions of device D3 that are 7, 8 and 9 layers thick, respectively. (f, g, h). Simulations for device D3.

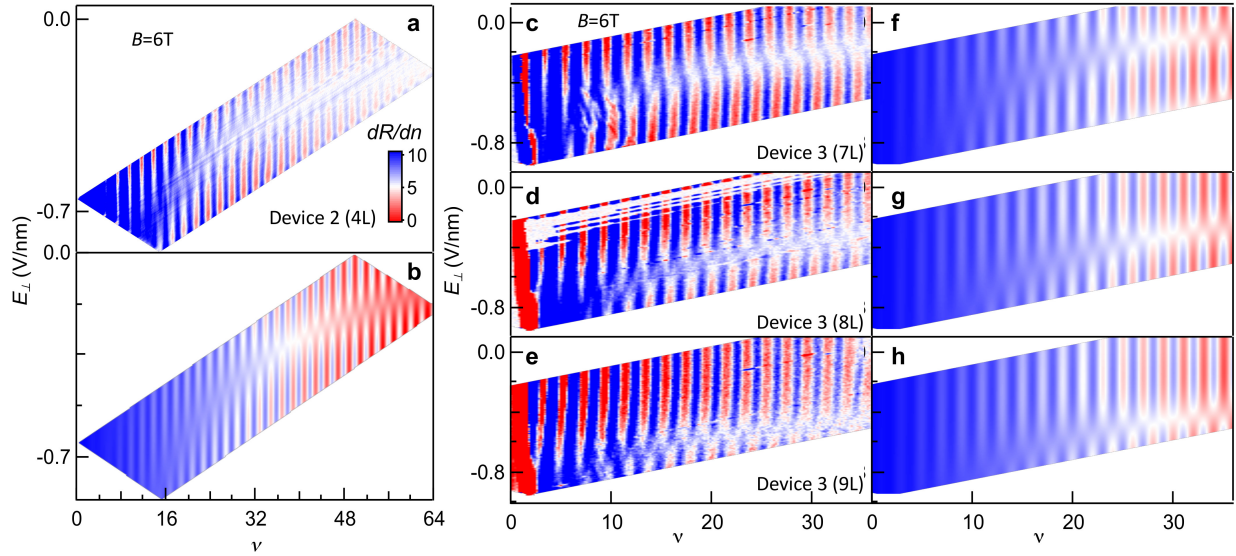

Fig. 4. Magneto-transport data and anomalously large SOC parameter. (a-b). $dR_{xx}/dn(n, E_\perp)$ and simulations at constant $B=10T$ for device D1. (c). Extracted Rashba parameter and its dependence on $E_\perp$ for different devices. (d). DFT calculations of Rashba parameter vs $E_{\perp,s}$ for different c-axis lattice constants. Here $E_{\perp,s}$ is the out-of-plane electric field used in calculations, which, due to screening, is significantly reduced from the experimental values calculated from external gate voltages.

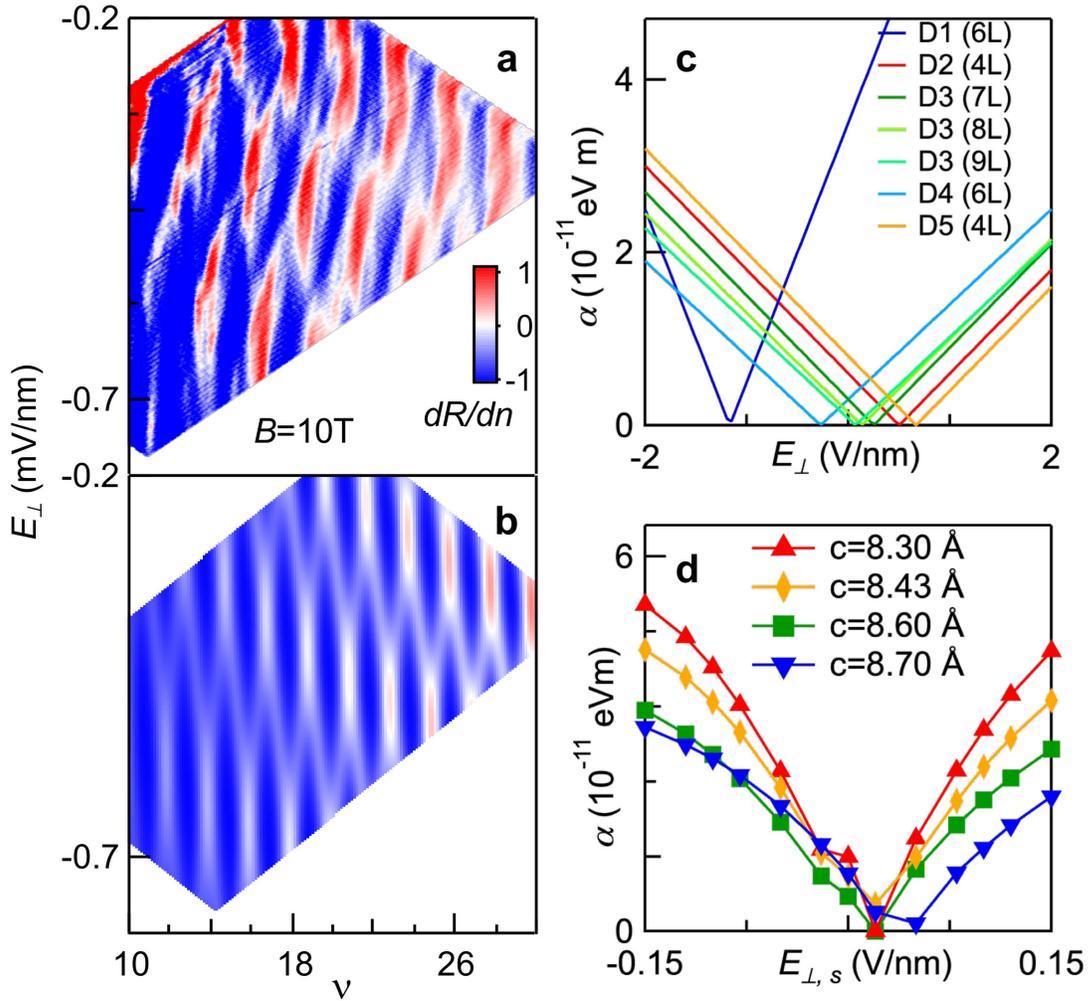